\documentclass[prl,aps,epsf]{revtex4}
\usepackage[english]{babel}
\usepackage{graphicx}
\usepackage{epsf}
\usepackage{pstricks}

\begin{document}

\hfill{CPHT-RR-036.0605}

\title{Search for isotensor exotic meson
and twist $4$ contribution to $\gamma^*\gamma\to\rho\rho$}

\author{I.V. Anikin $^{a,b,c}$, B. Pire$^b$ and O.V. Teryaev$^{c}$}
\affiliation{$^a$LPT, Universit{\'e} Paris-Sud, 91405 Orsay, France
             \footnote {Unit{\'e} mixte 8627 du CNRS} \\
             $^b$CPHT, {\'E}cole Polytechnique,
             91128 Palaiseau Cedex, France\footnote{Unit{\'e} mixte 7644 du CNRS} \\
             $^c$Bogoliubov Laboratory of Theoretical Physics,
             JINR, 141980 Dubna, Russia }
\vspace{1.5cm}

\begin{abstract}

\noindent
We present a theoretical estimate for the
cross-section of exclusive $\rho^+\rho^-$ and $\rho^0\rho^0$-meson
production in two photon collisions when one of the initial
photons is highly virtual. We focus on the discussion of the twist 4 contributions
which are related to the production of an exotic isospin $2$ resonance of two $\rho$
mesons.
Our analysis shows that the recent experimental data
obtained by the L3 Collaboration at LEP can be understood as a signal
for the existence of an exotic isotensor resonance with a mass around $1.5\, {\rm GeV}$.
\vspace{1pc}
\end{abstract}
\maketitle

\section{I. Introduction}
%\vspace{0.5cm}

\noindent
Exclusive reactions $\gamma^*\gamma\to A + B $ which may be accessed in
$e^+ e^-$ collisions have been shown \cite{DGPT} to have a partonic
interpretation in the kinematical region of large virtuality of one photon and
of small center of mass energy. The  scattering
amplitude factorizes in a long distance dominated object -- the
generalized distribution amplitude (GDA) -- and a short distance perturbatively
calculable scattering matrix. A phenomenological analysis of the $\pi \pi$
channel \cite{DGP} has shown
that precise experimental data could be collected at intense $e^+ e^-$
collider experiments such as BABAR and BELLE. Meanwhile, first data on the
$\rho^0 \rho^0$ channel at
LEP have been published \cite{L3Coll1} and analyzed \cite{APT}, showing
the compatibility of the QCD leading order analysis with experiment at quite modest
values of $Q^2$.

\noindent
In this paper, we focus on the comparison of processes $\gamma^* \gamma
\to \rho^0\rho^0$
and $\gamma^* \gamma \to \rho^+\rho^-$ in the context of searching
an exotic isospin $2$ resonance decaying in two $\rho$ mesons; such
channels have recently been studied at LEP by the L3 collaboration \cite{L3Coll1, L3Coll2}. A
related study for photoproduction \cite{Rosner}
raised the problem of $\rho^0\rho^0$ enhancement with respect to
$\rho^+\rho^-$ at low energies.
One of the solutions of this problem was based on the prediction \cite{Achasov0} and
further exploration \cite{Achasov} of
the possible existence
of isotensor state, whose interference with the isoscalar state is constructive for neutral mesons
and destructive for charged ones.
This option was also independently considered in \cite{Liu}.
The crucial property of such an exotic state is the absence of $\bar q q$ wave function at any
momentum resolution. In other words, quark-antiquark component is absent both in its non-relativistic
description and
at the level of the light-cone distribution amplitude. This is by no means common: for instance,
the $1^{-+}$ state which is a
quark-gluon hybrid at the non-relativistic level is described by a leading twist quark-antiquark
distribution amplitude \cite{AnHyb}.
Contrary to that, an isotensor state on the light cone corresponds to the twist $4$ or higher and
its  contribution is thus power suppressed at large $Q^2$.
This is supported by the mentioned L3 data, where
the high $Q^2$ ratio two of the cross sections of charged and neutral mesons production
points out an isoscalar state.

\noindent
We studied both perturbative and non-perturbative ingredients of QCD factorization for the
description of an isotensor state.
Namely, we calculated the twist $4$ coefficient function and extracted the non-perturbative
matrix elements from L3 data.
Our analysis is compatible with the existence of an isotensor exotic meson with a mass
around $1.5$  GeV.

\section{II. Amplitude of $\gamma^* \gamma \to \rho\rho$ process}

\noindent
The reaction which we study here is
$e(k)+e(l)\to e(k^{\prime})+e(l^{\prime})+\rho(p_1)+\rho(p_2)$,
where $\rho$ stands for the triplet $\rho$ mesons;
the initial  electron $e(k)$ radiates
a hard virtual photon with momentum $q=k-k^{\prime}$,
with $q^2=-Q^2$ quite large. This means that the scattered electron  $e(k^{\prime})$
is tagged.
To describe the given reaction, it is useful to consider the sub-process
$e(k)+\gamma(q^{\prime})\to e(k^{\prime})+\rho(p_1)+\rho(p_2)$.
Regarding the other photon momentum $q^{\prime}=l-l^{\prime}$,
we assume that, firstly, its momentum is almost collinear to the electron
momentum $l$ and, secondly, that $q^{\prime \, 2}$ is approximately
equal to zero, which is a usual approximation when the second lepton
is untagged.

\noindent
In two $\rho$ meson production, we are interested in the channel where
the resonance corresponds to the exotic isospin, {\it i.e} $I=2$, and usual
$J^{PC}$ quantum numbers. The $J^{PC}$ quantum numbers are not essential for our study.
Because the isospin $2$ has only a projection on the four quark correlators,
the study of mesons with the isospin $2$ can help to throw light upon the
four quark states. We thus, together with the mentioned reactions,
study the following processes:
$e(k)+e(l)\to e(k^{\prime})+e(l^{\prime})+ R(p)$
and $e(k)+\gamma(q^{\prime})\to e(k^{\prime})+R(p)$,
where meson $R(p)$ possesses isospin $I=2$.

\noindent
Considering the amplitude of the $\gamma^*\gamma$ subprocess, we write
\begin{eqnarray}
\label{amp01}
{\cal A}_{(i,j)}(\gamma\gamma^*\to\rho\rho)=\varepsilon^{\,\prime\,(i)}_{\mu} \varepsilon^{(j)}_{\nu}
\int d^4z_1 d^4z_2 \,e^{-iq^{\prime}\cdot z_1-iq\cdot z_2}
\langle \rho(p_1) \rho(p_2)| T\left[ J_{\mu}(z_1) J_{\nu}(z_2)\right] | 0 \rangle ,
\end{eqnarray}
where $J_\mu$ denotes the quark electromagnetic current
$J_\mu=\bar\psi {\cal Q}\gamma_\mu \psi$ with the charge matrix ${\cal Q}$ belonging to
$SU_F(2)$ group. The photon polarization vectors read
\begin{eqnarray}
\label{polvec}
\varepsilon^{\,\prime\,(\pm)}_{\mu}=
\left( 0,\frac{\mp 1}{\sqrt{2}},\frac{+i}{\sqrt{2}},0 \right), \quad
\varepsilon^{\,(\pm)}_{\mu}=
\left( 0,\frac{\mp 1}{\sqrt{2}},\frac{-i}{\sqrt{2}},0 \right), \quad
\varepsilon^{\,(0)}_{\mu}=
\left(\frac{|q|}{\sqrt{Q^2}},0,0,\frac{q_0}{\sqrt{Q^2}} \right),
\end{eqnarray}
for the real and virtual photons, respectively.
The coefficient functions of twist $2$ operators to Operator Product Expansion of currents product
in (\ref{amp01}) were discussed in detail in \cite{APT}, while
the contributions of new twist $4$ operators are described by coefficient functions calculated
long ago in \cite{JS} when considering the problem of twist $4$ corrections
to Deep Inelastic Scattering.

\noindent
Let us now turn on the flavour or isospin structure of the corresponding amplitudes.
The $\rho\rho$ state with $I=0$ can be projected on both the two and
four quark operators, while the state with $I=2$ on the four quark operator only.
Indeed, let us start from the consideration of the vacuum--to--$\rho\rho$ matrix element
in (\ref{amp01})
%in the zeroth order of the strong coupling constant. This case leads to the following correlator
\begin{eqnarray}
\label{cor2q}
\langle \rho^a \rho^b |
\bar\psi_f(0) \Gamma \psi_g(z) |0\rangle=
\delta^{ab} I_{fg}\Phi^{I=0}+
i\varepsilon^{abc}\tau^c_{fg}\Phi^{I=1},
\end{eqnarray}
where the quark fields are shown with free
flavour indices and $\Gamma$ stands for the corresponding $\gamma$-matrix.
The isoscalar and isovector GDA's $\Phi^{I}$ in (\ref{cor2q}) are well-known, see
for instance \cite{Diehlrep}. Note that, in (\ref{cor2q}), the correspondence between triplets
$\{\rho^1,\,\rho^2,\,\rho^3\}$ and $\{\rho^+,\,\rho^-,\,\rho^0\}$ is given by
the standard way.

\noindent
Moreover, for the coefficient function at higher order in the strong coupling constant,
the corresponding matrix element  gives us
\begin{eqnarray}
\label{cor4q}
\langle \rho^a \rho^b |
[\bar\psi_{f_1}(0) \Gamma_1 \psi_{g_1}(\eta)]
[\bar\psi_{f_2}(z) \Gamma_2 \psi_{g_2}(\xi)] |0\rangle.
\end{eqnarray}
Using the Clebsch-Gordan decomposition, we obtain
\begin{eqnarray}
\label{I0pr4q}
\biggl([\bar\psi_{f_1} \psi_{g_1}]\, [\bar\psi_{f_2} \psi_{g_2}]
\biggr)^{I=0,\,I_z=0}&\Rightarrow&
-\frac{1}{\sqrt{3}}\biggl[
\frac{1}{2} \tau^0_{f_1 g_1}\tau^0_{f_2 g_2} +
\tau^+_{f_1 g_1}\tau^-_{f_2 g_2} + \tau^-_{f_1 g_1}\tau^+_{f_2 g_2}
\biggr] \tilde\Phi^{I=0,\,I_z=0}
\end{eqnarray}
for the isospin $0$ and $I_z=0$
projection of the four quark operator in (\ref{cor4q}), and
\begin{eqnarray}
\label{I2pr4q}
\biggl([\bar\psi_{f_1} \psi_{g_1}]\, [\bar\psi_{f_2} \psi_{g_2}]
\biggr)^{I=2,\,I_z=0}&\Rightarrow&
\frac{1}{\sqrt{6}}\biggl[
\tau^0_{f_1 g_1}\tau^0_{f_2 g_2} -
\tau^+_{f_1 g_1}\tau^-_{f_2 g_2} - \tau^-_{f_1 g_1}\tau^+_{f_2 g_2}
\biggr] \tilde\Phi^{I=2,\,I_z=0}
\end{eqnarray}
for the isospin $2$ and $I_z=0$ projection of the four quark operator in (\ref{cor4q}).
The four quark GDA's $\tilde\Phi^{I,\,I_z=0}$ can be defined in an analogous manner as the two
quark GDA's.
Hence, one can see that the amplitudes (\ref{amp01})
for $\rho^0\rho^0$ and $\rho^+\rho^-$ productions can be written in the form of
the decomposition:
\begin{eqnarray}
\label{sumam}
{\cal A}_{(+,+)} = {\cal A}_{(+,+)\,2}^{I=0,\,I_z=0} +
{\cal A}_{(+,+)\,4}^{I=0,\,I_z=0} + {\cal A}_{(+,+)\,4}^{I=2,\,I_z=0},
\end{eqnarray}
where the subscripts $2$ and $4$ in the amplitudes imply that
the given amplitudes are associated with the two and four quark correlators, respectively.
The amplitudes corresponding to $\rho^+\rho^-$
production are not independent and can be expressed through the corresponding amplitudes
of $\rho^0\rho^0$ production. Indeed, one can derive the following relations:
\begin{eqnarray}
\label{rel1}
&&{\cal A}_{(+,+)\,k}^{I=0,\,I_z=0}(\gamma\gamma^*\to\rho^+\rho^-)=
{\cal A}_{(+,+)\,k}^{I=0,\,I_z=0}(\gamma\gamma^*\to\rho^0\rho^0)
\quad {\rm for}\,\,\, k=2,\,4
\nonumber\\
&&{\cal A}_{(+,+)\,4}^{I=2,\,I_z=0}(\gamma\gamma^*\to\rho^+\rho^-)=
-\frac{1}{2}{\cal A}_{(+,+)\,4}^{I=2,\,I_z=0}(\gamma\gamma^*\to\rho^0\rho^0).
\end{eqnarray}

\noindent
The amplitude of two $\rho$ meson production in two photon
collision can be also presented through a resonant intermediate state.
The vacuum  to $\rho\rho$ matrix element in the {\it r.h.s.} of (\ref{amp01}) can be traded for
\begin{eqnarray}
\label{meR}
\sum_{I=0,1,2}\langle \rho(p_1)\,\rho(p_2)| R^I(p)\rangle
\frac{1}{M^2_{R^I}-p^2-i\Gamma_{R^I} M_{R^I}}
\langle R^I(p)| T\left[ J_{\mu}(0) J_{\nu}(z)\right] |0\rangle.
\end{eqnarray}
where $R^I(p)$ is the resonance with three possible isospin $I=0,\,1,\,2$.
Note that, in our case, only isospin $0$ and $2$ cases are relevant due to
the positive $C$-parity of the initial and final states. The matrix element
$\langle\rho\,\rho| R^I\rangle$ defines the corresponding coupling constant of meson and
$\langle R^I| T\left[ J_{\mu}(0) J_{\nu}(z)\right] |0\rangle$ is considered up to the
second order of strong coupling constant $\alpha_S$, {\it i.e} this matrix element
is written as a sum of two- and four-quark correlators.

\section{III. Differential cross sections}

\noindent
Previously, the theoretical description of the experimental data collected for the $\rho^0\rho^0$
production has been performed in \cite{APT}.
Now, the subject of our study is the differential cross section corresponding to
both the $\rho^0\rho^0$ and $\rho^+\rho^-$ productions in the electron--positron collision.

\noindent
Using the equivalent photon approximation \cite{Budnev}
we find the expression for the corresponding cross section~:
\begin{eqnarray}
\label{xsec5}
\frac{d\sigma_{ee\to ee\rho\rho}}{dQ^2\,dW^2}=
\int..\int d{\rm cos}\theta\, d\phi\, dx_2
\frac{\alpha}{\pi} F_{WW}(x_2)
\frac{d\sigma_{e\gamma\to e\rho\rho}}
{ dQ^2 \,dW^2\, d{\rm cos}\theta\, d\phi},
\end{eqnarray}
where the usual Weizsacker-Williams function $F_{WW}$ is used.
In (\ref{xsec5}), the cross section for the subprocess reads
\begin{eqnarray}
\label{xsec6}
\frac{d\sigma_{e\gamma\to e\rho\rho}}
{dQ^2\, dW^2\, d{\rm cos}\theta\, d\phi}=
\frac{\alpha ^3}{16\pi}
\frac{\beta}{S_{e\gamma}^2}\,
\frac{1}{Q^2}
\Biggl(
1-\frac{2S_{e\gamma}(Q^2+W^2-S_{e\gamma})}
{(Q^2+W^2)^2}
\Biggr)
\left| A_{(+,+)} \right|^2
\end{eqnarray}
where the amplitude $A_{(+,+)}$ is defined by (\ref{sumam}).
For the case of $\rho^0\rho^0$ production, the cross section (\ref{xsec5}) takes the form:
\begin{eqnarray}
\label{xsec7}
&&\frac{d\sigma_{ee\to ee\rho^0\rho^0}}{dQ^2dW^2}=
\frac{100\alpha^4}{9} G(S_{ee},Q^2,W^2) \beta
\\
&&\Biggl(
\frac{\Gamma_{R^0} M_{R^0}}{\beta_0((M^2_{R^0}-W^2)^2+\Gamma_{R^0}^2M^2_{R^0})}
\biggl[ {\bf S}^{I=0,I_3=0}_2+\frac{\alpha_S(Q^2)M^2_{R^0}}{Q^2}{\bf S}^{I=0,I_3=0}_4\biggr]^2+
\Biggr.
\nonumber\\
\Biggl.
&&\frac{\Gamma_{R^2} M_{R^2}}{\beta_2((M^2_{R^2}-W^2)^2+\Gamma_{R^2}^2M^2_{R^2})}
\biggl[\frac{\alpha_S(Q^2)M^2_{R^2}}{Q^2}{\bf S}^{I=2,I_3=0}_4\biggr]^2+
\Biggr.
\nonumber\\
\Biggl.
&&2\sqrt{\frac{\Gamma_{R^0}\Gamma_{R^2} M_{R^0} M_{R^2}}{\beta_0 \beta_2}}
\frac{(M^2_{R^0}-W^2)(M^2_{R^2}-W^2)+(\Gamma_{R^0}M_{R^0})(\Gamma_{R^2}M_{R^2})}
{((M^2_{R^0}-W^2)^2+\Gamma_{R^0}^2M^2_{R^0})((M^2_{R^2}-W^2)^2+\Gamma_{R^2}^2M^2_{R^2})}\times
\Biggr.
\nonumber\\
\Biggl.
&&\biggl[ {\bf S}^{I=0,I_3=0}_2
+\frac{\alpha_S(Q^2)M^2_{R^0}}{Q^2}{\bf S}^{I=0,I_3=0}_4\biggr]
\frac{\alpha_S(Q^2)M^2_{R^2}}{Q^2}{\bf S}^{I=2,I_3=0}_4
\Biggr),
\nonumber
\end{eqnarray}
where $\Gamma_{R^I}$ stand for the total widths. The dimensionful structure constants
${\bf S}^{I,I_3=0}_4$ and ${\bf S}^{I=0,I_3=0}_2$ are related to the nonperturbative
vacuum--to--meson matrix elements.
The $\beta$--functions are also defines in the standard ways:
$\beta=\sqrt{1-4m_{\rho}^2/W^2}$ and $\beta_I=\sqrt{1-4m_{\rho}^2/M_{R^I}^2}$.
The function $G$ in (\ref{xsec7}) is equal to
\begin{eqnarray}
\label{Gfun}
G(S_{ee},Q^2,W^2)=\int\limits_{0}^{1}
dx_2 F_{WW}(x_{2})
\Biggl[
\frac{1}{ x_2^2 S_{ee}^2 Q^2}-
\frac{2}{ x_2 S_{ee} Q^2(Q^2+W^2)}+
\frac{2}{Q^2(Q^2+W^2)^2}
\Biggr].
\end{eqnarray}
The differential cross section corresponding to $\rho^+\rho^-$ production can be obtained using
(\ref{rel1}), we have
\begin{eqnarray}
\label{xsec8}
&&\frac{d\sigma_{ee\to ee\rho^+\rho^-}}{dQ^2dW^2}=
\frac{200\alpha^4}{9} G(S_{ee},Q^2,W^2) \beta
\\
&&\Biggl(
\frac{\Gamma_{R^0} M_{R^0}}{\beta_0((M^2_{R^0}-W^2)^2+\Gamma_{R^0}^2M^2_{R^0})}
\biggl[ {\bf S}^{I=0,I_3=0}_2+\frac{\alpha_S(Q^2)M^2_{R^0}}{Q^2}{\bf S}^{I=0,I_3=0}_4\biggr]^2+
\Biggr.
\nonumber\\
\Biggl.
&&\frac{1}{4}\frac{\Gamma_{R^2} M_{R^2}}{\beta_2((M^2_{R^2}-W^2)^2+\Gamma_{R^2}^2M^2_{R^2})}
\biggl[\frac{\alpha_S(Q^2)M^2_{R^2}}{Q^2}{\bf S}^{I=2,I_3=0}_4\biggr]^2-
\Biggr.
\nonumber\\
\Biggl.
&&\sqrt{\frac{\Gamma_{R^0}\Gamma_{R^2} M_{R^0} M_{R^2}}{\beta_0 \beta_2}}
\frac{(M^2_{R^0}-W^2)(M^2_{R^2}-W^2)+(\Gamma_{R^0}M_{R^0})(\Gamma_{R^2}M_{R^2})}
{((M^2_{R^0}-W^2)^2+\Gamma_{R^0}^2M^2_{R^0})((M^2_{R^2}-W^2)^2+\Gamma_{R^2}^2M^2_{R^2})}\times
\Biggr.
\nonumber\\
\Biggl.
&&\biggl[ {\bf S}^{I=0,I_3=0}_2
+\frac{\alpha_S(Q^2)M^2_{R^0}}{Q^2}{\bf S}^{I=0,I_3=0}_4\biggr]
\frac{\alpha_S(Q^2)M^2_{R^2}}{Q^2}{\bf S}^{I=2,I_3=0}_4
\Biggr),
\nonumber
\end{eqnarray}

\noindent
Note  that we have explicitly separated out, in (\ref{xsec7}) and (\ref{xsec8}),
the running coupling constant $\alpha_S(Q^2)$ which  appears in the twist $4$ terms.
Because of we will study the $Q^2$ dependence of the corresponding cross sections
at rather small values of $Q^2$, we use the Shirkov and Solovtsov's analytical
approach \cite{Shirkov} to determine the running coupling constant in the region
of small $Q^2$. Detailed discussion on different aspects of using the analytical
running coupling constant may be found in \cite{Bakulev, AnHyb} and references therein.

\section{IV. LEP data fitting}

\noindent
In the previous section we derived the differential cross sections
$d\sigma_{ee\to ee\rho\rho}/dQ^2\, dW^2$ for both the $\rho^0\rho^0$ and
$\rho^+\rho^-$  channels, based on the QCD analysis. These expressions contain a
number of unknown phenomenological parameters, which are intrinsically related to non perturbative
quantities encoded in the generalized distribution amplitudes.
One should now make a fit of these phenomenological parameters
in order to get a good description of experimental data.
The best values of the parameters can be found by the method of least squares, $\chi^2$-method,
which flows from the maximum likelihood theorem, but we postpone a comprehensive $\chi^2$-analysis
to a forthcoming more detailed paper. Here, we implement a naive fitting analysis to get an acceptable
agreement with the experimental data.
Thus, we have the following set of parameters for fitting:
\begin{eqnarray}
{\bf P}=\{M_{R^0},\,\Gamma_{R^0},\,
M_{R^2},\,\Gamma_{R^2},\,{\bf S}^{I=0,I_3=0}_2,\,{\bf S}^{I=0,I_3=0}_4,
\,{\bf S}^{I=2,I_3=0}_4\}.
\end{eqnarray}

\noindent
We start with the study of the $W$ dependence of the cross sections.
For this goal, following the papers \cite{L3Coll1, L3Coll2}, we determine the cross section
of process $ee\to ee\rho\rho$ normalized by the integrated luminosity function:
\begin{eqnarray}
\label{Wdep}
\sigma_{\gamma\gamma^*}(\langle W \rangle)=
%\frac{\int dQ^2 d\sigma_{ee\to ee\rho\rho}(Q^2,\langle W \rangle)}
%{\int dQ^2 {\cal L}(Q^2,\langle W \rangle)}=
\frac{\int dQ^2 {\cal L}(Q^2,\langle W \rangle)\sigma_{\gamma\gamma^*}(Q^2,\langle W \rangle)}
{\int dQ^2 {\cal L}(Q^2,\langle W \rangle)},
\end{eqnarray}
where the definition of the luminosity function ${\cal L}$ is taken from \cite{JF}. The value
$\langle W \rangle$ corresponds to the center of each bin, see \cite{L3Coll1, L3Coll2}.
Focussing first on the region of larger $Q^2$ we fit the parameters associated with the
dominant contribution which comes from the twist $2$ term amplitude, which is
associated with the non-exotic resonance (or background) with isospin $I=0$.
Generally speaking, there are many isoscalar resonances with masses in the region of
$1 - 3 $ GeV.
To include their total effect we introduce a mass and width for an "effective" isoscalar resonance.
We then  determine the values of the mass and width by fitting the data for the region
of larger $Q^2$ ({\it i.e.}, when $Q^2$ is  in the interval $1.2<Q^2<8.5$ GeV$^2$).
We thus can fix the parameters ${\bf S}^{I=0,I_3=0}_2$,
$M_{R^0}$ and $\Gamma_{R^0}$. Good agreement can be achieved with
$M_{R^0}=1.8\, {\rm GeV}$, $\Gamma_{R^0}=1.00\, {\rm GeV}$ and ${\bf S}^{I=0,I_3=0}_2$
within the interval $(0.12,\, 0.16)$. As can be expected the width of the effective isoscalar
"resonance" is fairly large. It means that we actually deal with a non-resonant background.

\noindent
Next, we fit the $W-$dependence of the cross section for small values of $Q^2$, {\it i.e.} $0.2<Q^2<0.85$
GeV$^2$.
In this region all twist contributions may be important.
We find that the experimental data can be described by the following choice
of the parameters: $M_{R^2}=1.5\, {\rm GeV}$, $\Gamma_{R^2}=0.4\, {\rm GeV}$ while
the parameters ${\bf S}^{I=0,I_3=0}_4$ and ${\bf S}^{I=2,I_3=0}_4$ are  in the
intervals $(0.002, \, 0.006)$ and $(0.012, \, 0.018)$, respectively.
\begin{figure}[htb]
$$\includegraphics[width=7cm]{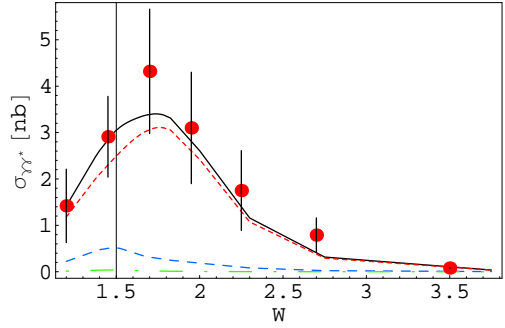}$$
\caption{$W$ dependence of the cross section $\sigma_{ee\to ee\rho^0\rho^0}$
normalized by the integrated luminosity function,
in the  $1.2<Q^2<8.5$ region. The short-dashed line corresponds
to the leading twist $2$ contribution;
the dash-dotted line to
the twist $4$ contribution; the middle-dashed line to the interference of twist $2$
and $4$ contributions.
The solid line corresponds to the sum of all contributions. }
\label{rho0rho02}
\end{figure}
\begin{figure}[htb]
$$\includegraphics[width=7cm]{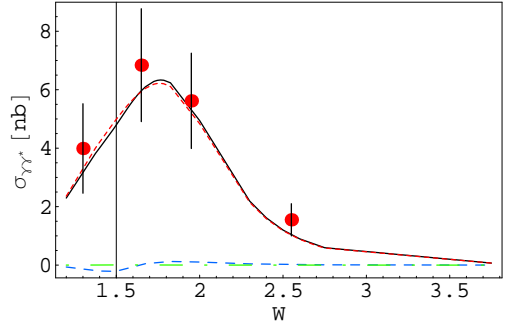}$$
\caption{$W$ dependence of the cross section $\sigma_{ee\to ee\rho^+\rho^-}$
with the same conventions as for Fig. \ref{rho0rho02} }
\label{rho+rho-2}
\end{figure}

\noindent
Further, we include in our analysis the $Q^2$ dependence of $\rho^0\rho^0$
and $\rho^+\rho^-$ production cross sections, {\it i.e.} $d\sigma_{ee\to ee\rho\rho}/dQ^2$,
which should fix the remaining arbitrariness of the parameters.
We  finally find that the best description of both $W$ and $Q^2$ dependence is reached at
\begin{eqnarray}
\label{sol_chi2}
&&M_{R^2}=1.5 \, {\rm GeV},
\quad \Gamma_{R^2}=0.4 \, {\rm GeV},
\nonumber\\
&&{\bf S}^{I=0,I_3=0}_2=0.12 \, {\rm GeV}, \quad
{\bf S}^{I=0,I_3=0}_4=0.006 \, {\rm GeV}, \quad
{\bf S}^{I=2,I_3=0}_4=0.018 \, {\rm GeV}.
\end{eqnarray}
Note that these rather small values of twist $4$ structure constants
${\bf S}_4$ compared to the twist $2$ structure constant ${\bf S}_2$
indicate that leading twist contribution dominate for the values $Q^2$ around or greater than
$1\, {\rm GeV}^2$. This should be compared with what was obtained in a particular renormalon model in \cite{AGP}.

\begin{figure}[htb]
$$\includegraphics[width=7cm]{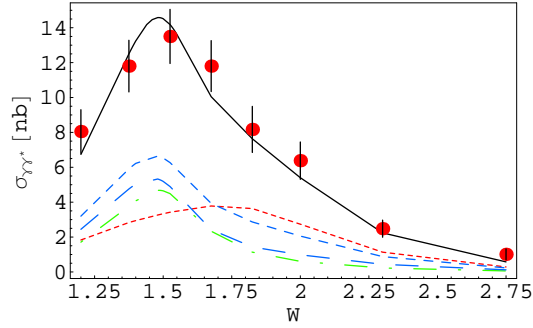}$$
\caption{$W$ dependence of the cross section $\sigma_{ee\to ee\rho^0\rho^0}$
normalized by the integrated luminosity function in the $0.2<Q^2<0.85$ region.
The short-dashed line corresponds
to the leading twist $2$ contribution;
the dash-dotted line to the twist $4$ contribution;
the middle-dashed line to
the contributions from the interference between twist $2$ and twist $4$ terms;
the long-dashed line to
the contribution from the interference between isoscalar and isotensor terms.
The solid line corresponds to the sum of all contributions. }
\label{rho0rho0}
\end{figure}

\begin{figure}[htb]
$$\includegraphics[width=7cm]{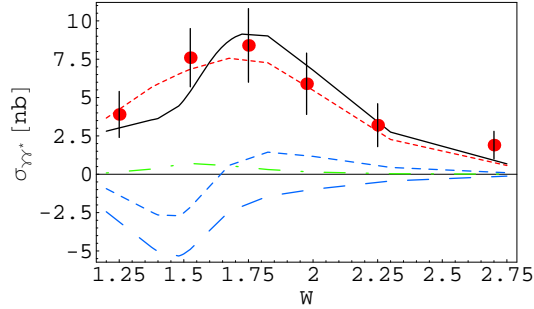}$$
\caption{same as Fig. \ref{rho0rho0} for the cross section $\sigma_{ee\to ee\rho^+\rho^-}$
in the $0.2<Q^2<0.85$ region.  }
\label{rho+rho-}
\end{figure}

\noindent
Our theoretical description of the LEP experimental data are presented on
 Figs. \ref{rho0rho02}--\ref{rhorho1}. The plots depicted on Figs. \ref{rho0rho02}--\ref{rho+rho-}
have the following notations: the short-dashed line corresponds
to the contribution coming from the leading twist term of (\ref{xsec7});
the dash-dotted line -- to
the contribution from the twist $4$ term of (\ref{xsec7}); the middle-dashed
line -- to
the contributions from the interference between twist $2$ and twist $4$ terms of (\ref{xsec7})
and (\ref{xsec8}); the long-dashed
line -- to
the contribution from the interference between isoscalar and isotensor terms.
Finally the solid line corresponds to the sum of all contributions.
On Fig. \ref{rhorho1}, we present the LEP data and our theoretical curves for
both the $\rho^0\rho^0$ and $\rho^+\rho^-$ production differential cross sections
as functions of $Q^2$.
The solid line on Fig. \ref{rhorho1} corresponds to the $\rho^0\rho^0$ differential cross section
while the dashed one -- to  the $\rho^+\rho^-$ differential cross section.

\section{V. Discussions and Conclusions}

\noindent
The fitting of LEP data based on the QCD factorization of the amplitude into a hard subprocess and
a generalized distribution amplitude thus allows us to claim  evidence of the
existence of an isospin $I=2$ exotic meson \cite{Achasov0,Achasov,Maiani} with a mass in the vicinity of $1.5\, {\rm GeV}$
and a width around $0.4\, {\rm GeV}$. The contributions of such an exotic meson in the two
 $\rho$ meson production cross sections (see, (\ref{xsec7}) and (\ref{xsec8})) are directly associated with
some twist $4$ terms that we have identified. At large $Q^2$, these twist $4$ contributions become
 negligible and the behaviours of the $\rho^0\rho^0$ and $\rho^+\rho^-$ cross sections are controlled
 by the leading twist $2$ contributions, see Fig. \ref{rho0rho02} and
\ref{rho+rho-2}.  Figs. \ref{rho0rho0} and \ref{rho+rho-} show the increasing role
of higher twist contributions when decreasing $Q^2$. Namely, the
interference between twist $2$ and $4$ amplitudes gives the dominant contributions to
$\rho^0\rho^0$ production in the lower $Q^2$ interval, and is thus responsible of the  $W$ dependence
of the cross section in these kinematics. In particular, in this interference term
the main contribution arises from the interference between isoscalar and isotensor structures,
see the long-dashed lines on Fig. \ref{rho0rho0} and \ref{rho+rho-}.

\begin{figure}[htb]
$$\includegraphics[width=7cm]{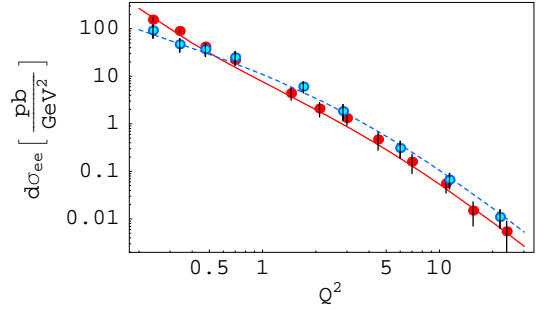}$$
\caption{The $Q^2$ dependence of the differential cross sections $d\sigma_{ee\to ee\rho^0\rho^0}/dQ^2$
and $d\sigma_{ee\to ee\rho^+\rho^-}/dQ^2$. The solid line corresponds
to the case of $\rho^0\rho^0$ production;
the dashed line to the case of $\rho^+\rho^-$ production.}
\label{rhorho1}
\end{figure}

\noindent
Analysing the $Q^2$ dependence, we can see that due to the presence of a twist $4$ amplitude
and its interference with the leading twist $2$ component, the $\rho^0\rho^0$ cross section at
 small $Q^2$ is a few times higher than the
$\rho^+\rho^-$ cross section, see Fig. \ref{rhorho1}.
While for the region of large $Q^2$ where any higher twist effects are
negligible the $\rho^0\rho^0$ cross section is less than the $\rho^+\rho^-$ cross section by
the  factor $2$, which is typical from an isosinglet channel (see also (\ref{xsec7}) and (\ref{xsec8})).

\noindent
The reaction $\gamma^* \gamma \to \rho \rho$ and its QCD analysis in the framework of
 Ref. \cite{DGPT} thus proves its efficiency to reveal facts on hadronic
physics which would remain quite difficult to explain in a quantitative way otherwise. The leading
twist dominance is seen to persist down to values of $Q^2$ around $1\, {\rm GeV}^2$. Other
aspects of QCD may be revealed in different kinematical regimes through the same reaction \cite{other}.
Its detailed experimental analysis at intense electron colliders within the BABAR and BELLE
experiments is thus extremely promising. Data at higher energies in a future linear
collider should also be foreseen.

\noindent
Note that the non-perturbative calculations of the relevant $I=2$ twist $4$
matrix elements also deserve special interest. In particular, one may follow the ideas developed for
pion distribution \cite{max} which allowed to relate
higher and lower twists in multicolour QCD. The generalization for the case of $\rho$ mesons, anticipated
by the authors  of \cite{max},  and use of crossing relations between various kinematical domains
provided by Radon transform technique \cite{radon} may allow to apply these result in the case under
consideration.

\noindent
In conclusion, let us stress that the L3 data allows to estimate the contribution of higher twist four quark
light cone distribution to the production amplitude of vector meson pairs. Our numerical analysis leads to a
rather small width for the corresponding resonant state, which is nothing else as an exotic four-quark
isotensor meson.
At the same time, a more elaborate experimental, theoretical and numerical analysis is required to
confirm, with better accuracy, the smallness of the width and the existence of an exotic meson.

\section{Acknowledgements}

\noindent
We are grateful to N.N.~Achasov, A.~Donnachie, J.~Field, K.~Freudenreich, M.~Kienzle, N.~Kivel, K.F.~Liu,
M.V.~Polyakov and I.~Vorobiev for useful discussions and correspondence.
O.V.T. is indebted to Theory Division of CERN and CPHT, {\'E}cole Polytechnique, for warm hospitality. I.V.A.
expresses gratitude to Theory Division of CERN and University of Geneva for
financial support of his visit.
This work has been supported  in part by RFFI Grant 03-02-16816.
I.V.A. thanks NATO for a grant.

\end{document}